\documentclass[aps,prl,twocolumn,psfig,showpacs,superscriptaddress]{revtex4}

\usepackage{amsfonts}
\usepackage{amsmath}
\usepackage{graphicx}
\usepackage{bm}
\usepackage{color}
\usepackage{amssymb}
\usepackage{ulem}
\usepackage{times}
\usepackage{dcolumn}
\usepackage{cases}
\usepackage{txfonts}

\begin{document}

\title{Identification of Degeneracy and Criticality of Two-Dimensional Statistical and Quantum Systems by the Boundary States of Tensor Networks}

\author{Shi-Ju Ran}
\affiliation{Theoretical Condensed Matter Physics and Computational Materials Physics Laboratory, School of Physics, University of Chinese Academy of Sciences, P. O. Box 4588, Beijing 100049, China}
\author{Cheng Peng}
\affiliation{Theoretical Condensed Matter Physics and Computational Materials Physics Laboratory, School of Physics, University of Chinese Academy of Sciences, P. O. Box 4588, Beijing 100049, China}
\author{Wei Li}
\affiliation{Department of Physics, Beihang University, Beijing 100191, China}
\author{Gang Su}
\email[Corresponding author. ]{Email: gsu@ucas.ac.cn}
\affiliation{Theoretical Condensed Matter Physics and Computational Materials Physics Laboratory, School of Physics, University of Chinese Academy of Sciences, P. O. Box 4588, Beijing 100049, China}

\begin{abstract}
  We propose a systematic scheme to reach the properties of two-dimensional (2D) statistical and quantum systems by studying the effective (1+1)-dimensional theory that is constructed from the tensor network representation. On on hand, we discover that the degeneracy of the 2D system can be determined by the purity of the boundary thermal state, which is the density operator of the effective theory at zero (effective) temperature. On the other hand, we find that the gapped (or critical) 2D system leads to a gapped (or critical) effective (1+1)-dimensional theory whose criticality can be accessed by the entanglement entropy $S$ of its ground state dubbed as boundary pure state. We also uncover that for the critical systems, $S$ obeys the same logarithmic law as that found in the critical 1D quantum chains, which reads $S = (\kappa c/6)\log_2 D + const.$, with $c$ the central charge and $\kappa$ a constant related to the scaling property of the correlation length $\xi$ as $\xi \sim D^{\kappa}$. Such a scaling law presents an efficient way to characterize the critical universality class of the original 2D systems. An important implication of our work is that many well-established theories for 1D quantum chains become available for studying 2D systems with the help of the proposed lower dimensional correspondence.
\end{abstract}


\maketitle

\textit{Introduction.}--- Lots of efforts have been made on exploring novel properties of quantum many-body systems in two dimensions (2D), where the fruitful geometries of the 2D lattices and the strong competition between the quantum fluctuation and magnetic ordering provide a fertile ground for various exotic phenomena. For instance, the 2D frustrated Heisenberg models such as the kagom\'{e} antiferromagnet \cite{Kagome1} were shown to be good candidates for realizing quantum spin liquids (QSL's) \cite{QSL}. It is of basic scientific interest to study the elusive properties of the QSL's such as topological orders \cite{TopoOrder}, fractional excitations \cite{FractionalE} and criticality \cite{QPT}.

However, many important issues are hard to access in 2D. Because the bipartition of a 2D system suffers much more complexity than that in one dimension (1D), quantities which involves the subregion size scaling, e.g. the central charge which characterizes the critical universality class \cite{CFT,CFT_Ent} and the topological entanglement entropy which can be used to identify topological orders \cite{Stopo}, are extremely difficult to reach. A usual way to make a compromise is to calculate on a finite torus \cite{Kagome1,Critical2D}, which suffers the finite size effect. Meanwhile, classical simulations are inefficient on critical systems even in 1D \cite{Entangle1}, which makes the criticality in 2D more elusive.


The boundary theories that are recently proposed based on the tensor network (TN) representation \cite{BoundaryH0,BoundaryH1,BoundaryH2} present us a new clue to solve problems that cannot be easily handled with conventional methods. The tensor network state (TNS, also termed as projected entangled pair state) \cite{TPS1st,PEPS} gives a natural representation for a quantum state, where the local tensors act as the projectors that map the physical degrees of freedom to virtual ones carrying the entanglement. The inner product of a TNS and its copy $\langle \psi_{TNS} |\psi_{TNS} \rangle$ maps a quantum many-body state to its classical correspondence in the same dimension. The resulting classical model can be further mapped onto an effective 1D quantum model. Some theories based on this scheme have been proposed to tackle difficult problems, e.g. identifying the topological orders \cite{BoundaryH1} and simulating low-energy excitations \cite{BoundaryH2} of 2D quantum systems, etc.

In this work, we propose a correspondence from the TN representation of a statistical system or a quantum state in 2D to an effective (1+1)-dimensional theory $\mathcal{H}$. Different from the existing theories that are based on the reduced density matrix, $\mathcal{H}$ is constructed straightforwardly by the transfer matrix of the TN. We suggest that the properties of the original 2D system can be efficiently reached by investigating this lower dimensional correspondence $\mathcal{H}$.

First, the boundary thermal state (BTS) is introduced as the thermal state of $\mathcal{H}$ at zero effective temperature, which is obtained by the linearized tensor renormalization group (LTRG) method \cite{LTRG}. Note the LTRG was proposed to obtain the thermodynamic properties of 1D quantum systems. We find the purity of the BTS a robust quantity to determine the degeneracy of the 2D system, i.e. the BTS is pure if the corresponding parent Hamiltonian \cite{ParentH} is non-degenerate, and is mixed if the parent Hamiltonian is degenerate.

Second, the boundary pure state (BPS) defined as the ground state of the (1+1)-dimension theory can be obtained by the infinite time-evolving block decimation (iTEBD) \cite{iTEBD,Canonical} method. Note that the iTEBD was suggested for calculating the ground states of 1D quantum chains. We discover that the entanglement entropy ($S$) of the BPS can be utilized to detect the criticality of the system. For gapped systems, the BPS is found to bear only a finite $S$, while for critical systems, $S$ increases with the dimension $D$ of the BPS, obeying the same logarithmic law as that found in the critical 1D quantum chains \cite{Entangle1}, which reads
\begin{equation}
 S = \frac{\kappa c}{6}\log_2 D + const.,
 \label{eq-ScalingS}
\end{equation}
with $c$ the central charge \cite{CFT} and $\kappa$ a constant related to the scaling of the correlation length $\xi$ defined as $\xi \sim D^{\kappa}$ \cite{Entangle1}. This gives an efficient way to determine the critical universality class of the 2D statistical and quantum systems through its (1+1)-dimensional correspondence, while the difficulties in scaling against the subregion length in 2D are avoided.

To examine our proposal, we perform exact deductions and numerical simulations on several celebrated examples including the 2D Ising model at and away from the critical temperature, the Greenberger-Horne-Zeilinger (GHZ) state (gapped and non-topological) \cite{GHZ}, the Z$_2$ state (gapped and topological) \cite{StringNet}, the nearest-neighbor resonating valence bond (NNRVB) state on kagom\'{e} (gapped and topological) and honeycomb (critical and topological) lattices \cite{RVBPEPS}.

\textit{Correspondence from a 2D TN to a (1+1)-dimensional theory.}--- A planar TN is defined by the contraction of tensors
\begin{equation}
 Z = tTr(\prod_{j} T^{[j]}_{u_jl_jd_jr_j}),
 \label{eq-TN}
\end{equation}
where $T^{[j]}_{u_jl_jd_jr_j}$ is the local tensors, $j$ runs over all tensors and $tTr$ is the trace of all common bonds. Eq. (\ref{eq-TN}) plays a central role in the TN scheme since the calculations of many physical quantities (e.g. the partition function of the 2D classical models \cite{TRG} or the inner product of two quantum states \cite{PEPS}) is equivalent to computing such a TN contraction. Here we take the square TN as an example [Fig. \ref{fig-TN} (a)].

Let us first explain how to apply the LTRG and iTEBD methods to calculating Eq. (\ref{eq-TN}). Without losing the generality, we presume the TN satisfies the translational invariance, i.e. $T^{[j]}_{u_jl_jd_jr_j} = T_{uldr}$. We start the contraction with the original local tensor $P^{(0)}_{uldr} = T_{uldr}$. For the $t$th step of the contraction, the effective tensor $P^{(t)}$ is updated by contracting one original tensor $T_{uldr}$ to $P^{(t-1)}$ as $P^{(t)}_{ul''d'r''} = \sum_{d u'} T_{uldr} P^{(t-1)}_{u'l'd'r'} \delta_{du'}$ with $l''=(l,l')$ and $r''=(r,r')$, as shown in Fig. \ref{fig-TN} (a). Such a contraction scheme is equivalent to applying the LTRG method to contract a 2D planar TN instead of performing the imaginary time evolution of the 1D quantum systems in its original proposal \cite{LTRG}. The TN is contracted to an matrix product operator (MPO) \cite{MPO} formed by $P^{(t)}$ and the entanglement spectrum $\lambda^O$ residing on the virtual bonds. The dimensions of $l''$ and $r''$ of $P$ are bounded in the same way as that the LTRG bounds the dimensions of the MPO, i.e. the truncations are implemented according to the entanglement spectrum $\lambda^O$ after canonicalizing the MPO. The local canonical conditions of an MPO, which is similar to the canonicalization of an matrix product state (MPS) \cite{Canonical}, are shown in Fig. \ref{fig-TN} (b). Specifically speaking, an MPO is canonical when the spectrum on any bond is the singular value spectrum between the left/right subregions.

At the $t$th step, $P^{(t)}$ represents efficiently a tensor stripe containing $t+1$ original tensors [dash squares in Fig. \ref{fig-TN} (a)], and the MPO represents the TN with $t+1$ layers. As $t \rightarrow \infty$, the MPO converges to the fixed point, i.e. $P_{ul''d'r''} = \sum_{d u'} T_{uldr} P_{u'l'd'r'} \delta_{du'}$. The fixed point MPO dubbed as the BTS represents effectively the whole TN.

Such a contraction of a planar TN can also be done accurately using the MPS with the iTEBD algorithm \cite{Canonical,PEPS}. After making a sufficiently large number of contractions, one obtains the converged MPS with its local tensor and entanglement spectrum denoted by $A$ and $\lambda^S$. We dub such an MPS that is pure as the BPS of the TN.

\begin{figure}[tbp]
\includegraphics[angle=0,width=0.85\linewidth]{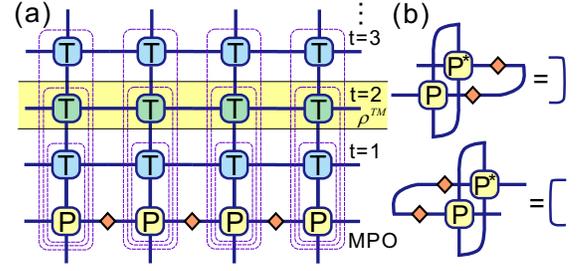}
  \caption{(Color online) (a) The LTRG scheme on a square TN formed by the local tensor $T$ and its copies. The contraction of the TN is performed linearly with the help of an MPO formed by local tensors $P$ and spectrum $\lambda^O$. At the $t$th renormalization step, each local tensor effectively represents a tensor strip (the dashed circles) containing ($t+1$) original tensors. The transfer matrix $\rho^{TM}$ of the TN is defined as the tensor stripe along the horizontal direction. (b) The left and right canonical conditions for the MPO.}
\label{fig-TN}
\end{figure}

The contractions of the TN through the LTRG and iTEBD are actually equivalent to the imaginary time evolution of a (1+1)-dimensional theory. To see this, we introduce the transfer operator $\rho^{TM}$ of the TN defined by the infinite tensor stripe along the direction perpendicular to the contraction direction [the yellow shadow in Fig. \ref{fig-TN} (a)]. The Hamiltonian $\mathcal{H}$ of the (1+1)-dimensional theory then can defined as $e^{-\mathcal{H}} = \rho^{TM} $. Here, we require the TN to have spatial inversion symmetry so that $\mathcal{H}$ is guaranteed to be hermitian.

In the LTRG, we readily have the MPO at the $t$th step as $\rho_{MPO}^{(t)} = e^{-\mathcal{K} \mathcal{H}}$ with $\mathcal{K}$ the effective inverse temperature satisfying $\mathcal{K} = t+1$. In the iTEBD, contracting the $t$th layer of the tensors in the TN to the MPS is actually equivalent to evolving the MPS along the effective imaginary time direction as $|MPS \rangle_t = e^{-\mathcal{H}} |MPS \rangle_{t-1} = e^{-\mathcal{K} \mathcal{H}} |MPS\rangle_0$ with $|MPS\rangle_0$ the initial MPS. The BPS defined as the fixed point MPS is the ground state of $\mathcal{H}$, fulfilling $e^{-\mathcal{H}} |BPS\rangle = \mathcal{C} |BPS\rangle$ ($\mathcal{C}$ is a constant) which is just the fixed point condition.

\textit{Purity of the boundary thermal state.}--- Now we show that the degeneracy of the ground state(s) of $\mathcal{H}$ can be detected by the purity of the BTS, and suggest that such a degeneracy can identify the degeneracy of the original 2D system. When $\rho^{TM}$ is non-degenerate, the BPS $|\psi_{BPS}\rangle$ is simply the ground of $\mathcal{H}$, and the BTS $\hat{\rho}_{BTS}$ gives a pure state that is the outer product of $|\psi_{BPS}\rangle$ and its copy, say $\hat{\rho}_{BTS} = |\psi_{BPS}\rangle \langle \psi_{BPS}|$. Consequently, the entanglement spectra $\lambda^O$ and $\lambda^S$ of the BTS and BPS have a simple outer-product relation $\lambda^O=\lambda^S \otimes \lambda^S$.


When $\mathcal{H}$ is degenerate, the BTS is no longer pure, i.e. it cannot be decomposed into an outer product form. The BTS can be formally written in a mixed thermal state $\hat{\rho}_{BTS} = \sum_{i=1}^{\chi} \eta_i |\phi^{i} \rangle \langle \phi^{i}|$ with $|\phi^i \rangle$ the $i$th degenerate eigenvector, $\eta$ the thermal probability distribution and $\chi$ the degeneracy. Because the degenerate eigenvectors of $\rho^{TM}$ correspond to a same eigenvalue, the BTS given by LTRG (which is essentially a power method) is expected to be the maximal mixture of the degenerate states, i.e. $\eta_i = 1/\chi$. Meanwhile, it is widely accepted that the BPS $|\psi_{BPS}\rangle$ by iTEBD favors the minimally entangled state among all combinations of the degenerate eigenvectors. In this case, the entanglement spectra of the BTS and BPS do not have a simple outer product relation. Thus we have a way to identify the purity of the BTS by monitoring the spectrum difference $\varepsilon = |\lambda^O - \lambda^S \bigotimes \lambda^S|$.


We observe that the degeneracy of $\mathcal{H}$ is in accordance with the degeneracy of the original 2D system. We shall remark that by the degeneracy of a quantum state, we mean the degeneracy of its parent Hamiltonian \cite{ParentH} (note a statistical system can be mapped to a quantum state \cite{PEPSCritical}). Interestingly, the BTS gives a mixed state whether the parent Hamiltonian is non-topological (e.g. the GHZ state) or topological (e.g. the Z$_2$ topological state). Thus the purity of the BTS is robust to detect the degeneracy of the system under interest.

\textit{Entanglement scaling of the boundary pure state.}--- Many remarkable works have been done with the entanglement of the MPS to study the properties of the ground states of 1D quantum models: the subregion length-dependence of the entanglement of a critical state of a spin chain is analogous to that of the entropy in conformal field theories \cite{Entangle1}; the entanglement exhibits the correlation properties and the critical exponents \cite{Entangle2} of the system as well as the topological properties, such as the symmetry-protected topological orders \cite{EntSpect1} and edge excitations \cite{EntSpect2}. Through our correspondence, these achievements become available for investigating 2D systems.

We suggest that the original 2D system shares the same criticality with its lower dimensional correspondence $\mathcal{H}$. Speaking in detail, for a gapped 2D system, the $\mathcal{H}$ is gapped where the BPS (the ground state of $\mathcal{H}$) possesses only finite entanglement entropy from the known conclusions in 1D quantum chains. For a critical 2D system, the $\mathcal{H}$ is critical, where the entanglement entropy defined as $S(\lambda^S) = -\sum_i (\lambda^S_i)^2 \log_2 (\lambda^S_i)^2$ of the BPS obeys a logarithmic scaling law given by Eq. (\ref{eq-ScalingS}).


To study the criticality of (infinite) strongly correlated systems directly in 2D is a extremely hard task both theoretically and numerically: (a) the conventional subregion scaling for the central charge $S(\lambda^S) = (c/3)\log_2 L + \zeta'$ \cite{CFT_Ent} is difficult to implement in 2D since the bipartition is essentially different from that in 1D; (b) any finite-entanglement approximation of a critical system gives an effective gapped model whose correlation length is finite. We speculate that the central charge of $\mathcal{H}$ which can be efficiently obtained \cite{Entangle1} can characterize the critical universality class of the original 2D system. Thus the difficulties encountered in 2D can be avoided.

\textit{Two exactly contractible states.}--- We give the exact deduction on two non-trivial states, the GHZ \cite{GHZ} and Z$_2$ topological state \cite{StringNet} to examine our proposal on detecting degeneracy. The GHZ state is a topological-trivial state whose parent Hamiltonian has two degenerate ground states. The Z$_2$ state is the exact ground state of the toric-code model which is gapped and possesses nontrivial topological degeneracy \cite{QCMP}. 

The local tensor of the TN in Eq. (\ref{eq-TN}) of the GHZ and Z$_2$ states can be written in a unified form $T_{uldr} = \sum_{\mu\mu_1\mu_2\mu_3\mu_4=0}^{1} \Lambda_{\mu} I_{\mu \mu_1 \mu_2 \mu_3 \mu_4} U_{u\mu_1} U_{l\mu_2} U_{d\mu_3} U_{r\mu_4}$ with $\Lambda$ the rank spectrum. We have $\Lambda_0 = \Lambda_1 = 1/\sqrt{2}$ and $U$ a ($2\times 2$) unitary matrix \cite{SM}. $I$ is the super-diagonal tensor satisfying $I_{\mu\mu_1\cdots}=1$ if $\mu=\mu_1=\cdots$ or $I_{\mu\mu_1\cdots}=0$ otherwise. The matrix $U$ is the rotation matrix satisfying $U_{0,0}=-U_{1,1}=\cos \theta$ and $U_{0,1}=U_{0,1}=\sin \theta$ with $\theta$ the parameter. By taking $\theta = 0$ and $\pi/4$, $U$ becomes the identical and Fourier matrix, which gives the TN for the GHZ and Z$_2$ states, respectively.

The BTS of the GHZ and Z$_2$ TN can also be expressed in a unified form with the local tensor $P_{uldr} = \sum_{\mu_1\mu_2} I_{\mu_1l\mu_2r} U_{u\mu_1} U_{d\mu_2}$ and the spectrum $\lambda^O=\Lambda$ \cite{SM}. It is easy to check that the BTS is canonical. Obviously, such a BTS is mixed and possesses a finite bond dimension $D=2$ with the spectrum $\lambda^O_0 = \lambda^O_1 = 1/\sqrt{2}$. For the BPS, the MPS converges to an unstable fixed point $A_{ulr} = \sum_{\mu} I_{\mu lr} U_{u\mu}$ and $\lambda^S=\Lambda$ only by choosing $A^{(0)} = I$ as the initial MPS. We have the entanglement entropy $S(\lambda^S) = \log_2 2 =1$ for both the GHZ and Z$_2$ states. Otherwise the fixed point flows to the stable fixed point with zero entanglement of the MPS.

It is not surprising to have only $D=2$ of the BTS and BPS for the GHZ state, as it is just the superposition of two classical ferromagnetic states. In contrast, for the Z$_2$ state, the quantum entanglement entropy (of the physical degrees of freedom) obeys the area law and increases unboundedly as $\mathcal{S}=\alpha L - \gamma$ with $L$ the boundary length of the subsystem and $\gamma$ the topological entanglement entropy \cite{Stopo}. Consequently, a large bond dimension $\chi \sim 2^{\mathcal{S}} \sim 2^{\alpha L}$ is expected to capture the quantum entanglement directly in 2D. Amazingly, the ground state of its lower dimensional correspondence only has a small dimension $D=2$ with a finite entanglement entropy.

We show that the absence of the physical degeneracy leads to a pure BTS. If we destroy the degeneracy of the ground states of the parent Hamiltonian, we have, equivalently, a shift of $\Lambda$ as $\Lambda_2 = \epsilon \Lambda_1$ ($0<\epsilon <1$). Then for the $t$th step of renormalization, the canonical spectrum of the MPO satisfies $\lambda^O_2 = \epsilon^{(t+1)} \lambda^O_1$. As $t\rightarrow \infty$, $\lambda^O_2$ vanishes and the BTS becomes a pure state. This picture also holds for the Z$_2$ state.

\textit{The 2D Ising model.}--- The partition function of the 2D antiferromagnetic (AF) Ising model on square lattice can be written as the contraction of a TN as Eq. (\ref{eq-TN}). The local tensor reads $T_{uldr} = \exp\{-[s_us_l+s_ls_d+s_ds_r+s_rs_u + h_s (s_u-s_l+s_d-s_r)]/\mathrm{T}\}$, where $s_u$, $s_l$, $s_d$ and $s_r$ are the four spins in a plaquette, $h_s$ is the staggered magnetic field and $\mathrm{T}$ is the temperature.

\begin{figure}[tbp]
\includegraphics[angle=0,width=1\linewidth]{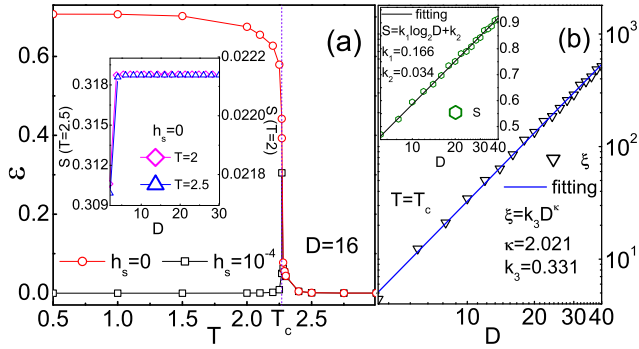}
  \caption{(Color online) (a) The temperature-dependence of $\varepsilon = |\lambda^O - \lambda^S \otimes \lambda^S|$ of the 2D Ising model with and without a small staggered magnetic filed $h_s$. The inset of (a) gives the entanglement entropy $S$ of the BTS versus its bond dimension $D$ at $\mathrm{T} = 2$, $2.5$. (b)  At $\mathrm{T} = \mathrm{T}_c$, the correlation length $\xi$ scales with $D$ as $\xi \sim D^{\kappa}$ with $\kappa=2.021$. The inset of (b) presents that $S$ satisfies $S = k_1\log_2 D + const.$, with $k_1=0.166$. According to Eq. (\ref{eq-ScalingS}), we have the central charge $c=0.493$.}
\label{fig-Ising}
\end{figure}

We study the purity of the BTS at different temperatures by numerically calculating $\varepsilon = |\lambda^O - \lambda^S \otimes \lambda^S|$ shown in Fig. \ref{fig-Ising} (a). For $\mathrm{T} < \mathrm{T}_c$ with $\mathrm{T}_c$ the critical temperature, we calculated $\varepsilon$ with (or without) a small staggered field $h_s= 10^{-4}$ where the Z$_2$ symmetry breaking is triggered (or conserved) in the LTRG scheme. We have $\lambda^O_0=1$, $\lambda^O_{s\geq1}=0$ and $\lambda^S=\lambda^O$ with $h_s= 10^{-4}$, which means the BTS is pure ($\varepsilon=0$). For $h_s=0$, the system contains two degenerate states (Z$_2$ symmetry is conserved) and the BTS flows to a GHZ-like mixed state for $\mathrm{T} < \mathrm{T}^c$ [$\lambda^O_0=\lambda^O_1 \simeq 1/\sqrt{2}$, $\lambda^O_{s\geq2} < O(10^{-2})$, $\lambda^S_0 \simeq 1$, $\lambda^S_{s\geq1} < O(10^{-2})$] and $\varepsilon$ is non-zero.

For $\mathrm{T} > \mathrm{T}^c$, the MPO flows to a trivial disordered state with or without a staggered field, and the BTS gives a pure state with vanishing $\varepsilon$. Consequently, the separating point of the two curves of $\varepsilon$ gives the critical temperature accurately, where we have $\mathrm{T}_c = 2.27$, in comparison to the exact critical temperature $\mathrm{T}_c = 2/\ln(1+\sqrt{2}) \simeq 2.269$. One can also see that at both sides of the critical temperature, the entropy $S(\lambda^S)$ saturates to a finite value when the bond dimension $D$ increases, as shown in the inset of Fig. \ref{fig-Ising} (a).

At $\mathrm{T}=\mathrm{T}_c$, $\varepsilon$ is non-zero indicating the BTS is a mixed state. We remark that the purity of the BTS at the critical point is not robust as an MPO with finite $D$ cannot accurately give true BTS of the TN. Fig. \ref{fig-Ising} (b) gives the correlation length $\xi$ versus $D$, which satisfies $\xi \sim D^{\kappa}$ with $\kappa=2.021$ indicating the (1+1)-dimensional theory $\mathcal{H}$ is critical \cite{Entangle1}. Such a criticality is also supported by the logarithmic scaling law of $S$ of the BPS shown in the inset of Fig. \ref{fig-Ising} (b). Then we accurately have the central charge $c=0.493$ from Eq. (\ref{eq-ScalingS}), while the exact result is $c=1/2$ from the conformal field theory \cite{CFT}.


\textit{The topological RVB states.}--- The TNS representations of the NNRVB states are given in Ref. [\onlinecite{PEPSCritical}]. It is known that the NNRVB, which is topological, is critical on bipartite lattices \cite{RVBCritic} but gapped on non-bipartite lattices \cite{RVBPEPS}.

\begin{figure}[tbp]
\includegraphics[angle=0,width=1\linewidth]{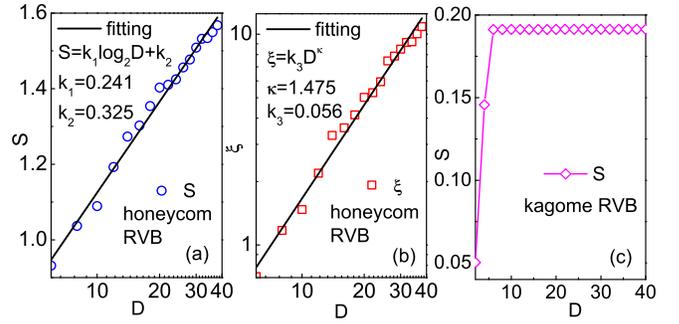}
  \caption{(Color online) The scaling of (a) the entanglement entropy $S$, (b) the correlation length $\xi$ of the honeycomb NNRVB state and (c) the entanglement entropy $S$ of the kagom\'{e} NNRVB state against the bond dimension $D$ of the BPS. For the honeycomb NNRVB state which is critical, $S$ obey a logarithmic scaling law $S = k_1\log_2 D + const.$ with $k_1=0.241$ and $\xi$ satisfies $\xi \sim D^{\kappa}$ with $\kappa=1.475$, which gives the central charge $c=0.98 \approx 1$ from Eq. (\ref{eq-ScalingS}). For the kagom\'{e} NNRVB state, $S$ saturates with $D$ showing such a state is gapped.}
\label{fig-RVB}
\end{figure}

Fig. \ref{fig-RVB} (a) shows that the entanglement entropy $S$ of the BPS of the honeycomb RVB state increases with $D$ in a logarithmic way as Eq. (\ref{eq-ScalingS}) with the coefficient $(\kappa c)/6=0.241$, and we have $\kappa=1.475$ with the scaling of the correlation length $\xi$ shown in Fig. \ref{fig-RVB} (b). Then we have the central charge of the NNRVB on honeycomb lattice $c=0.98\approx 1$.

For the kagom\'{e} RVB state which is gapped, we find $\varepsilon \simeq 0.1$, indicating that the BTS is mixed. In Fig. \ref{fig-RVB} (c), we observe $S$ saturates at about $S \simeq 0.19$, evidencing that our theory can identify whether a 2D quantum state is gapped or gapless by the criticality of the (1+1)-dimensional correspondence. We do not see the degeneracy in the entanglement spectrum of the BPS, which we have however seen in the Z$_2$ state. The reason may be that the corresponding symmetry is not well protected during the contraction procedure \cite{SPTNR}.

\textit{Conclusion.}--- In summary, we propose that the degeneracy and criticality of a 2D system can be efficiently identified by the purity of the BTS and the entanglement entropy of the BPS, both of which are associated to a lower dimensional correspondence $\mathcal{H}$. We suggest that the central charge of $\mathcal{H}$ which is reached by the finite dimension scaling can be used to characterize the criticality of the original 2D systems. Exact deductions and numerical calculations on different kinds of examples give strong supports to our proposal, where we uncover that the NNRVB state on honeycomb lattice bears the central charge $c\approx1$.

Our proposal could be readily extended to study other critical phenomena and critical quantum states such as the entropy-driven phase transitions \cite{PartialOrder}, critical quantum lattice-gas states \cite{LatGasState} and chiral spin liquids \cite{PEPSChiral}, as well as the variational ground states obtained by the TN-based algorithms.


This work is supported in part by the MOST of China (Grant No. 2012CB932900 and No. 2013CB933401), the Strategic Priority Research Program of the Chinese Academy of Sciences (Grant No. XDB07010100), and the NSFC (Grant No. 11474279).

\bigskip
\bigskip
\bigskip
\setcounter{figure}{0}
\setcounter{equation}{0}
\renewcommand{\thefigure}{S\arabic{figure}}

\large{\large{\large{\textbf{Supplemental Material}}}}
\maketitle

\bigskip
\noindent \textbf{I. Tensor networks of the GHZ and Z$_2$ states}
\bigskip

The GHZ \cite{GHZ} state is a highly entangled state which is defined as $|\psi_{GHZ}\rangle = \frac{1}{\sqrt{2}} (\prod_{i=1}^N |0\rangle_i + \prod_{j=1}^N |1\rangle_j )$ and has been introduced as a source state for quantum computations \cite{GHZ_QC}. The GHZ state in form of a square TNS is
\begin{eqnarray}
 | \psi_{GHZ} \rangle = \sum_{\{a,s\}} \prod_{n} I_{a_{n,1}a_{n,2}a_{n,3}a_{n,4}} \prod_{j} I_{s_j a_j a'_j} | s_j \rangle,
\label{eq-PEPSGHZ}
\end{eqnarray}
where $| s_j \rangle$ with $s_j =0$, $1$ denotes the up or down eigenstate of the $j$th spin and $\{a\}$ are the ancillary indices. $I$ is the super-diagonal tensors defined as
\begin{eqnarray}
 I_{a_1a_2 \cdots a_n}&=&
 \left\{
 \begin{array}{lll}
  1, \ \ a_1=a_2=\cdots=a_n. \\
  0, \ \ otherwise. 
 \end{array}
 \right.
\label{eq-DiagT}
\end{eqnarray}

The Z$_2$ topological state is the ground state of Z$_2$ Hamiltonian $\hat{H}_{Z_2} = -U \sum_I \prod_{i \in legs \ of \ I} \hat{\sigma}^x_i - t \sum_p \prod_{j \in edges \ of \ p} \hat{\sigma}^z_j $ ($t \gg U$) with $\hat{\sigma}^{\alpha}$ the Pauli operator \cite{StringNet}. It is a topologically ordered state with long range entanglement. Its TNS representation is
\begin{eqnarray}
 | \psi_{Z_2} \rangle = \sum_{\{a,s\}} \prod_{n} Q_{a_{n,1}a_{n,2}a_{n,3}a_{n,4}} \prod_{j} I_{s_j a_j a'_j} | s_j \rangle,
\label{eq-PEPSZ2}
\end{eqnarray}
where $Q$ is
\begin{eqnarray}
 Q_{a_{n,1}a_{n,2}a_{n,3}a_{n,4}}&=&
 \left\{
 \begin{array}{lll}
  1, \ \sum_{\alpha=1}^4 a_{n,\alpha}=even. \\
  0, \ otherwise. 
 \end{array}
 \right.
\label{eq-TensorsZ2}
\end{eqnarray}

\begin{figure}[htbp]
  \includegraphics[angle=0,width=1\linewidth]{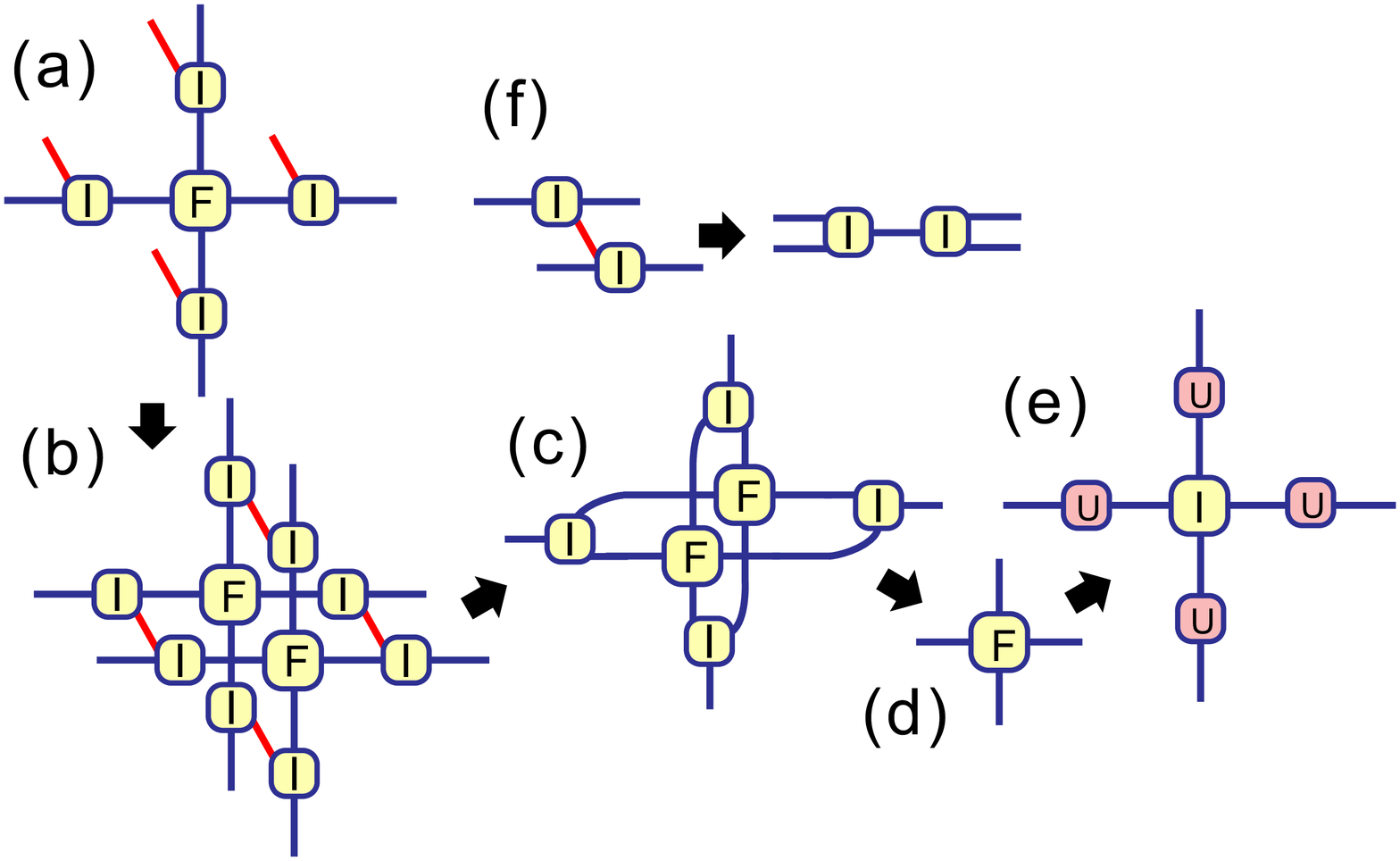}
  \caption{(Color online) (a) The unified TNS representation of the GHZ and Z$_2$ state [Eq. (\ref{eq-TNS})]. (b) The sketch of the inner product TN $Z = \langle \psi|\psi \rangle$. (c) The TN $Z$ can be transformed with Eq. (\ref{eq-TNZ}) into a TN formed by the local tensor $F$ shown in (d). (e) $F$ can be decomposed by the orthogonal tensor decomposition with Eq. (\ref{eq-TNStheta}). (f) The eigenvalue decomposition of the tensor $I$ [Eq. (\ref{eq-eigdec})].}
  \label{fig-Z2}
\end{figure}

One can see that the TNS of such two states can be written in a unified form
\begin{eqnarray}
| \psi \rangle= \sum_{\{a,s\}} \prod_{n} F^{(n)}_{a_{n,1}a_{n,2}a_{n,3}a_{n,4}} \prod_{j} I^{(j)}_{s_j a_j a'_j} | s_j \rangle,
\label{eq-TNS}
\end{eqnarray}
where $F^{(n)}$ is a tensor defined on the $n$th vertex of the network and $I^{(j)}$ is a super-diagonal tensor [Eq. (\ref{eq-DiagT})] on the $j$th edge. See FIG. \ref{fig-Z2} (a). The tensor $I^{(j)}$ is actually a projector which projects the ancillary indices represented by $a_j$ and $a'_j$ to the physical index $s_j$.

With the given TNS, a calculation of concerned quantity, such as $Z = \langle \psi|\psi \rangle$ or $\langle \hat{O} \rangle = \langle \psi|\hat{O}|\psi \rangle / Z$ with $\hat{O}$ a quantum operator, becomes the contraction of a corresponding TN. For the TN of $Z$, the local tensor satisfies [FIG. \ref{fig-Z2} (c)]
\begin{equation}
\begin{split}
 T^{(n)}_{g_{n,1}g_{n,2}g_{n,3}g_{n,4}} =  \sum_{a,b} F^{(n)}_{a_{n,1}a_{n,2}a_{n,3}a_{n,4}} F^{(n)*}_{b_{n,1}b_{n,2}b_{n,3}b_{n,4}} \\ I^{(n,1)}_{a_{n,1}b_{n,1}g_{n,1}}
 I^{(n,2)}_{a_{n,2}b_{n,2}g_{n,2}}I^{(n,3)}_{a_{n,3}b_{n,3}g_{n,3}} I^{(n,4)}_{a_{n,4}b_{n,4}g_{n,4}},
\end{split}
\label{eq-TNZ}
\end{equation}
where $I^{(n,j)}_{a_{n,j}b_{n,j}g_{n,j}} (j=1,2,3,4)$ is obtained by the eigenvalue decomposition [FIG. \ref{fig-Z2} (f)]
\begin{eqnarray}
 \sum_{s_{n,j}} I^{(n,j)}_{s_{n,j}a_{n,j}a'_{n,j}} I^{(n,j)}_{s_{n,j}b_{n,j}b'_{n,j}} = \sum_{g_{n,j}} I^{(n,j)}_{a_{n,j}b_{n,j}g_{n,j}} I^{(n,j)}_{a'_{n,j}b'_{n,j}g_{n,j}}.
\label{eq-eigdec}
\end{eqnarray}

When taking the local tensor $F^{(n)}$ as the one in Eq. (\ref{eq-PEPSGHZ}) or Eq. (\ref{eq-PEPSZ2}), one readily has $T^{(n)} = F^{(n)}$ in Eq. (\ref{eq-TNZ}) by substitutions (or according to the tensor fusion algebra \cite{AlgeTN}). Meanwhile, we found that $T^{(n)}$ can be decomposed by orthogonal tensor decomposition \cite{OTD} as
\begin{eqnarray}
 T^{(n)}_{a_{n,1}a_{n,2}a_{n,3}a_{n,4}} = \sum_{r=0}^{1} \Lambda^{(n)}_r U_{a_{n,1}r} U_{a_{n,2}r} U_{a_{n,3}r} U_{a_{n,4}r},
\label{eq-TNStheta}
\end{eqnarray}
The matrix $U$ for GHZ and Z$_2$ TN is actually the rotation matrix
\begin{eqnarray}
 &U =
 \left[
\begin{array}{cc}
 \cos(\theta) & \sin(\theta) \\
 \sin(\theta) & -\cos(\theta) \\
\end{array}
\right]
\label{eq-Vmatrix}
\end{eqnarray}
with $\theta$ the parameter. $U$ is orthogonal which satisfies
\begin{eqnarray}
 I_{rr'} = \sum_{a} U_{ar} U_{ar'}.
\label{eq-Identity}
\end{eqnarray}
By taking $\Lambda^{(n)} = \Lambda = [1/\sqrt{2},1/\sqrt{2}]$ in Eq. (\ref{eq-TNStheta}) and $\theta=0$ in Eq. (\ref{eq-Vmatrix}), one obtains the GHZ state with Eq. (\ref{eq-TNS}) and the corresponding local tensor of $Z$ with Eq. (\ref{eq-TNStheta}). By taking $\Lambda = [1/\sqrt{2},1/\sqrt{2}]$ and $\theta=\pi/4$, one has the Z$_2$ topologically ordered state and the local tensor of $Z$. Comparing these two states, both of them are highly entangled, while the GHZ state was introduced for quantum teleportation and the Z$_2$ state was revealed to bear a non-trivial topological entanglement. But against expectations, such two states belong to the same class characterized by only one parameter $\theta$.

Note that any translational invariant TN can be exactly contracted when the local tenor bears the same form as the that of the GHZ/Z$_2$ TN shown above, which actually gives the orthogonal tensor decomposition \cite{OTD}; the contraction properties of the Z$_2$ state has been studied from a different perspective using the Hopf algebra \cite{AlgeTN}.

\bigskip
\noindent \textbf{II. The fixed-point MPO of the GHZ/Z$_2$ tensor networks}
\bigskip

\begin{figure}[tbp]
\includegraphics[angle=0,width=1\linewidth]{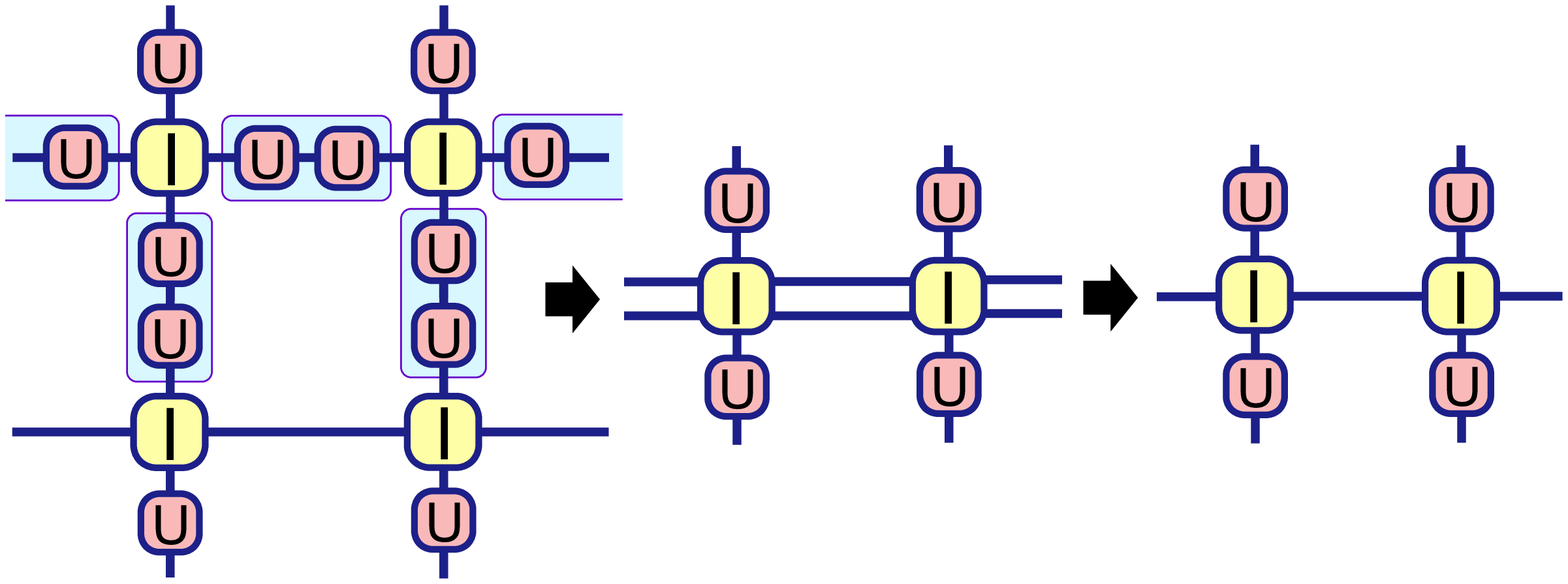}
  \caption{(Color online) The sketch that shows the fixed point MPO. For the first arrow, we apply the orthogonality $UU^T=I$, and for the second arrow, we apply the equations $\sum_{\mu} I_{\mu \mu_1 \cdots} I_{\mu \mu_1 \cdots} = I_{\mu_1 \cdots \mu_2 \cdots} $ and $\sum_{\mu\mu'} I_{\mu \mu' \cdots} I_{\mu \mu' \cdots} = \sum_{\mu} I_{\mu \cdots} I_{\mu \cdots} $.}
\label{fig-Exact}
\end{figure}

FIG. \ref{fig-Exact} illustrates the proof of the fixed point of the GHZ and Z$_2$ TN with LTRG. The local tensor of the MPO satisfies
\begin{eqnarray}
 P_{a_{1}a_{2}a_{3}a_{4}} = \sum_{r=0}^{1} \Lambda_r U_{a_{1}r} I_{a_{2}r} U_{a_{3}r} I_{a_{4}r},
\label{eq-LocalT}
\end{eqnarray}
and the entanglement spectrum is $\lambda^O=\Lambda$. During the contraction, the adjacent matrices $U$'s vanish according to Eq. (\ref{eq-Identity}), and by applying $\sum_{\mu} I_{\mu \mu_1 \cdots} I_{\mu \mu_1 \cdots} = I_{\mu_1 \cdots \mu_2 \cdots} $ and $\sum_{\mu\mu'} I_{\mu \mu' \cdots} I_{\mu \mu' \cdots} = \sum_{\mu} I_{\mu \cdots} I_{\mu \cdots} $, the resulting MPO is exactly the same as the one before the contraction. This proof holds for both GHZ and Z$_2$ TNs by taking $\theta=0$ and $\pi/4$ in $U$ [Eq. (\ref{eq-Vmatrix})].


\begin{thebibliography}{99}

\bibitem{Kagome1} S. Yan, D. Huse, and S. R. White, Science \textbf{332}, 1173-1176 (2011).
\bibitem{QSL} L. Balents, Nature \textbf{464}, 199 (2010).
\bibitem{TopoOrder} X. G. Wen, Phys. Rev. B \textbf{40}, 7387 (1989).
\bibitem{FractionalE} R. B. Laughlin, Rev. Mod. Phys. \textbf{71}, 863-874 (1999).
\bibitem{QPT} S. Sachdev, \textit{Quantum Phase Transitions}, 2nd ed. (Cambridge University Press, Cambridge, 2011).
\bibitem{CFT} P. Di Francesco, P. Mathieu, and D. S\'{e}n\'{e}chal, \textit{Conformal Field Theory} (Springer, Heidelberg, 1999).
\bibitem{CFT_Ent} C. Holzhey, F. Larsen, and F. Wilczek, Nucl. Phys. B \textbf{424}, 443 (1994).
\bibitem{Stopo} A. Kitaev and J. Preskill, Phys. Rev. Lett. \textbf{96}, 110404 (2006); M. Levin and X. G. Wen, \textit{ibid}, 110405 (2006).
\bibitem{Critical2D} Such as S. Inglis and R. G. Melko, New J. Phys. \textbf{15}, 073048 (2013); J. Helmes, and S. Wessel, Phys. Rev. B \textbf{89}, 245120 (2014).
\bibitem{Entangle1} L. Tagliacozzo, Thiago. R. de Oliveira, S. Iblisdir, and J. I. Latorre, Phys. Rev. B \textbf{78}, 024410 (2008); F. Pollmann, S. Mukerjee, A. M. Turner, and J. E. Moore, Phys. Rev. Lett. \textbf{102}, 255701 (2009).
\bibitem{BoundaryH0} J. I. Cirac, D. Poilblanc, N. Schuch, and F. Verstraete, Phys. Rev. B \textbf{83}, 245134 (2011).
\bibitem{BoundaryH1} N. Schuch, D. Poilblanc, J. I. Cirac, and D. P\'{e}rez-Garc\'{i}a, Phys. Rev. Lett. \textbf{111}, 090501 (2013).
\bibitem{BoundaryH2} S. Yang, L. Lehman, D. Poilblanc, K. Van Acoleyen, F. Verstraete, J. I. Cirac, and N. Schuch, Phys. Rev. Lett. \textbf{112}, 036402 (2014).
\bibitem{TPS1st} H. Niggemann, A. Kl\"{u}mper and J. Zittartz, Z. Phys. B \textbf{104}, 103 (1997); Eur. Phys. J. B \textbf{13}, 15 (2000).
\bibitem{PEPS} F. Verstraete and J. I. Cirac, arXiv:cond-mat/0407066; J. Jordan, R. Or\'{u}s, G. Vidal, F. Verstraete, and J. I. Cirac, Phys. Rev. Lett. \textbf{101}, 250602 (2008).
\bibitem{LTRG} W. Li, S. J. Ran, S. S. Gong, Y. Zhao, B. Xi, F. Ye, and G. Su, Phys. Rev. Lett. \textbf{106}, 127202 (2011); S. J. Ran, W. Li, B. Xi, Z. Zhang, and G. Su, Phys. Rev. B \textbf{86}, 134429 (2012).
\bibitem{ParentH} N. Schuch, I. Cirac, and D. P\'{e}rez-Garc\'{i}a, Ann. Phys. 325, 2153 (2010).
\bibitem{iTEBD} G. Vidal, Phys. Rev. Lett. \textbf{91}, 147902 (2003); Phys. Rev. Lett. \textbf{98}, 070201 (2007).
\bibitem{Canonical} R. Or\'{u}s and G. Vidal, Phys. Rev. B \textbf{78}, 155117 (2008).
\bibitem{GHZ} W. D\"{u}r, G. Vidal, and J. I. Cirac, Phys. Rev. A \textbf{62}, 062314 (2000).
\bibitem{StringNet} Z. C. Gu, M. Levin, B. Swingle, and X. G. Wen, Phys. Rev. B \textbf{79}, 085118 (2009); O. Buerschaper, M. Aguado, and G. Vidal, \textit{ibid}. 085119 (2009).
\bibitem{RVBPEPS} D. Poilblanc, N. Schuch, D. P\'{e}rez-Garc\'{i}a, and J. I. Cirac, Phys. Rev. B \textbf{86}, 014404 (2012); N. Schuch, D. Poilblanc, J. I. Cirac, and D. P\'{e}rez-Garc\'{i}a, \textit{ibid}, 115108 (2012).
\bibitem{TRG} M. Levin and C. P. Nave, Phys. Rev. Lett. \textbf{99}, 120601 (2007).
\bibitem{MPO} B. Pirvu, V. Murg, J. I. Cirac and F. Verstraete, New J. Phys. \textbf{12} 025012 (2010).
\bibitem{PEPSCritical} F. Verstraete, M. M. Wolf, D. Perez-Garcia, and J. I. Cirac, Phys. Rev. Lett. \textbf{96}, 220601 (2006).
\bibitem{Entangle2} G. Vidal, J. I. Latorre, E. Rico, and A. Kitaev, Phys. Rev. Lett. \textbf{90}, 227902 (2003).
\bibitem{EntSpect1} F. Pollmann, A. M. Turner, E. Berg and M. Oshikawa, Phys. Rev. B \textbf{81}, 064439 (2010).
\bibitem{EntSpect2} X. L. Qi, H. Katsura, and Andreas W. W. Ludwig, Phys. Rev. Lett. \textbf{108}, 196402 (2012).
\bibitem{QCMP} A. Y. Kitaev, Ann. Phys. \textbf{303}, 2-30 (2003).
\bibitem{SM} See more mathematical details of the constructions of the TN's in the supplemental material.
\bibitem{RVBCritic} A. F. Albuquerque and F. Alet, Phys. Rev. B \textbf{82}, 180408 (2010); H. Ju, A. B. Kallin, P. Fendley, M. B. Hastings, and R. G. Melko, \textit{ibid} \textbf{85}, 165121 (2012).
\bibitem{SPTNR} A. Weichselbaum, Anna. of Phys. \textbf{327}, 2972¨C3047 (2012); C.-Y. Huang, X. Chen, and F.-L. Lin, Phys. Rev. B \textbf{88}, 205124 (2013).
\bibitem{PartialOrder} Q. N. Chen, M. P. Qin, J. Chen, Z. C. Wei, H. H. Zhao, B. Normand, and T. Xiang, Phys. Rev. Lett. \textbf{107}, 165701 (2011).
\bibitem{LatGasState} S. Tanaka, R. Tamura and H. Katsura, Phys. Rev. A \textbf{86}, 032326 (2012).
\bibitem{PEPSChiral} T. B. Wahl, H.-H. Tu, N. Schuch, and J. I. Cirac, Phys. Rev. Lett. \textbf{111}, 236805 (2013).
\end{thebibliography}

\begin{thebibliography}{99}
\bibitem{GHZ} W. D\"{u}r, G. Vidal, and J. I. Cirac, Phys. Rev. A \textbf{62}, 062314 (2000).
\bibitem{GHZ_QC} D. Gottesman and I. L. Chuang, nature \textbf{402}, 390-393 (1999).
\bibitem{StringNet} Z. C. Gu, M. Levin, B. Swingle, and X. G. Wen, Phys. Rev. B \textbf{79}, 085118 (2009); O. Buerschaper, M. Aguado, and G. Vidal, \textit{ibid}. 085119 (2009).
\bibitem{AlgeTN} S. J. Denny, J. D. Biamonte, D. Jaksch and S. R. Clark, J. Phys. A: Math. Theor. \textbf{45} 015309 (2012).
\bibitem{OTD} J. Chen and Y. Saad, SIAM. J. Matrix Anal. $\&$ Appl. \textbf{30}, 1709-1734 (2009).
\end{thebibliography}
\end{document}